# The internal energies of Heisenberg magnetic systems

Huai-Yu Wang[1,*] and Liang-Jun Zhai[1] and Meichun Qian[2]

1. *Department of Physics, Tsinghua University, Beijing 100084, China*
2. *Department of Physics, Virginia Commonwealth University, Richmond, Virginia 23284, United States*

*wanghuaiyu@mail.tsinghua.edu.cn

**Abstract**

The internal energies, including transverse and longitudinal parts, of quantum Heisenberg systems for arbitrary spin S are investigated by the double-time Green's function method. The expressions for ferromagnetic (FM) and antiferromagnetic (AFM) systems are derived when one-component of magnetization is considered with the higher order longitudinal correlation functions being carefully treated. An unexpected result is that around the order–disorder transition points the neighboring spins in a FM (AFM) system are more likely longitudinally antiparallel (parallel) than parallel (antiparallel) to each other for S$\leq$3/2 in spite of the FM (AFM) exchange between the spins. This is attributed to the strong quantum fluctuation of the systems with small S values. We also present the expressions of the internal energies of FM systems when the three-component of magnetizations are considered.

## I. Introduction

The quantum Heisenberg model has been studied enduringly. The double-time Green's function method [1], being applicable in the whole temperature range, has been employed to solve the model over half a century [2-19]. For a long time, the magnetization along the *z* direction was calculated, with the assumption that the components other than this direction were zero [2-11]. Since 2000, a skill has been developed to calculate all the three components of the magnetization [12-19]. In calculation of the magnetizations, the equation of motion (EOM) of the Green's functions is applied, and the higher-order Green's functions are usually decoupled to the lower-order ones in terms of the well-known Tyablikov decoupling [2], also called the random phase approximation (RPA).

It is generally believed that evaluation of the magnetizations under the RPA is quite reasonable. However, the internal energies obtained up to now have not been satisfactory. The internal energies of antiferromagnetic (AFM) lattices at temperature close to zero was discussed [3]. A viewpoint was that it was better to go beyond the RPA in order to achieve satisfactory internal energies [18]. That is to say, higher-order Green's functions have to be solved. However, it is very difficult to do so. There has not been much work [20–26] attempting to solve the higher-order Green's functions and they were usually limited to the low-dimensional lattices and the lowest spin quantum number $S=1/2$. Even for the low-dimensional systems, it was difficult to deal with the cases with higher spin quantum numbers. The only instance of dealing with the higher S values was confined in finite lattice site systems [25]. A remarkable progress was the calculation of the internal energies of ferromagnetic (FM) lattices above the Curie point by means of the higher-order Green's functions [27]. There was one common feature in the work presented in Refs. [20–27]: the higher-order Green's functions were constructed in the cases where the magnetization was zero.

To sum up, the evaluation of the internal energy of the Heisenberg systems when the magnetization was not zero by means of the Green's function method has seldom been there to see. We believe that under the RPA, it is possible to obtain as good as possible expressions for the internal energy applicable to any S value for nonzero magnetization.

The internal energy of a Heisenberg magnetic system mainly includes two parts, the transverse correlation energy (TCE) and longitudinal correlation energy (LCE), as defined in Eqs. (3) and (4) below. The former is easily calculated by means of the well-known spectral theorem without any further approximation [5,18,28]. Hereafter, when no further approximation is made in giving an expression of the energy, we say the expression is precise. In this sense, the expression of the TCE is precise. The LCE, however, can be dealt with precisely only in the case of $S=1/2$ and 1 [18,28,29]. For higher S values, the treatment of this part is troublesome. At first thought, the following approximation can be taken [5]:

$$\langle S_i^z S_j^z \rangle \approx \langle S_i^z \rangle \langle S_j^z \rangle, \quad i \neq j \tag{1}$$

where the subscripts label the lattice sites. In previous work, we also employed this approximation [30]. However, this approximation is obviously too rough. A good approximation of the longitudinal correlation function valid for any S value and temperature is still desirable. In this paper, we will present satisfactory expressions of the internal energies for some magnetic systems.

## II. One-component magnetization: Ferromagnetic systems

The Hamiltonian reads

$$H = -\frac{1}{2} J \sum_{i,j} S_i^- S_j^+ - \frac{1}{2} J \sum_{i,j} S_i^z S_j^z - K_z \sum_i (S_i^z)^2 - b_z \sum_i S_i^z \tag{2}$$

Throughout this paper, we consider the nearest neighbor exchanges, and the lower case English letters label lattice sites. In Eq. (2), $J>0$. The first two terms reflect the transverse and longitudinal correlations between the neighboring magnetic moments, respectively. At finite temperature, any moment has an instant orientation along the directions other than the *z* direction, which is embodied in the transverse correlation function. The third term is the single-ion anisotropy along the *z* direction that forces the spontaneous magnetization along this direction. The fourth term is the Zeeman energy when an external magnetic field along the *z* direction $b_z$ is applied. For the sake of convenience, we hereafter denote $S_p=S(S+1)$.

The internal energy is defined as the statistical average of the Hamiltonian per site, $U_{IN} = \langle H \rangle / N$, where *N* is the total site number in the system. Thus, the first terms of the internal energy are written as

$$U_{TC} = -\frac{1}{2} J \sum_j \langle S_0^- S_j^+ \rangle \tag{3}$$

and

$$U_{LC} = -\frac{1}{2} J \sum_j \langle S_0^z S_j^z \rangle, \tag{4}$$

and are termed as transverse and longitudinal correlation energies, respectively. The subscript 0 means the origin and the summations are taken over its nearest neighbors.

As has been mentioned, the first rough approximation made for the longitudinal correlation energy was Eq. (1). This was plausible when checking its value at two special temperatures, zero and the Curie point. At *T*=0, the internal energy, with the absence of the external field and anisotropy, is

$$U_{IN}(T=0) = U_{LC}(T=0) = -\frac{1}{2} J(0) S^2 \tag{5}$$

where we have defined $J(0) = c_1 J$ with the $c_1$ being the nearest neighbor number. The quantity $J(0)$ is in fact the case of taking the wave vector **k**=0 in the Fourier component of the exchange parameter $J(\mathbf{k}) = J \sum_a e^{i\mathbf{k}\cdot\mathbf{a}}$, where the summation is over the nearest neighbors of the origin.

Eq. (5) is the rigorous ground-state energy of a FM system. At the Curie point TC, the spontaneous magnetization becomes zero. So it seems plausible that $U_{LC}(T=T_C) = 0$. However, the analysis is not reliable. At *T*=0, Eq. (5) happens to be correct for FM lattices because the spontaneous magnetization is along the z direction and there is no transverse correlation between neighboring spins. In the case of an AFM system, even at zero temperature, the neighboring spins are not rigorously antiparallel to each other, and there is the transverse correlation effect. At this point Eq. (1) exposes its drawback. At order–disorder transition temperature such as TC (TN) for FM (AFM) lattices, although the spontaneous magnetization becomes zero, the LCE may not be zero due to the existence of the short-range correlation effect [27,31–33]. Hence, a smart treatment of this energy is required. In the following, we make use of the Green's function method to derive the expressions of the energies.

The double-time Green's function is defined as $G_{ij}(t,t') = \langle\langle A_i(t); B_j(t') \rangle\rangle$ where the two operators in the present section are chosen as $A = S^+$, $B(u) = e^{uS^z}S^-$.

Note that there is a parameter $u$ in the operator $B$. In applying the EOM method, the first job is to reckon the commutator of an operator $S_k^+$ and the Hamiltonian:

$$[S_k^+, H] = -J\sum_j (S_k^z S_j^+ - S_k^+ S_j^z) + K_2(S_k^z S_k^+ + S_k^+ S_k^z) + b_z S_k^+. \tag{6}$$

Then the higher order Green's functions are decoupled by the RPA. Subsequently, the Fourier component of $G_{ij}(t,t')$, denoted as $g(\mathbf{k},\omega)$, is solved:

$$g(\mathbf{k},\omega) = \frac{[A, B(u)]}{\omega - \omega(\mathbf{k})}. \tag{7}$$

The dispersion relation is $\omega(\mathbf{k}) = (|J(0)| + K_z C)(1 - \gamma_{kC})\langle S^z\rangle + b_z$, where the notation $\gamma_k$ is defined as $\gamma_{kC} = J(\mathbf{k})/(|J(0)| + K_z C)$. The coefficient $C$ in Eq. (12) comes from the Anderson-Callen version of the decoupling concerning the single-ion term [6].

By means of the well-known spectral theorem, we obtain the evaluation of the correlation function of the two operators:

$$\langle B_m(u,t') A_l(t)\rangle = \langle [A, B(u)]\rangle \sum_k \frac{e^{-i\omega(\mathbf{k})(t-t')}}{e^{\beta\omega(\mathbf{k})} - 1} e^{-i\mathbf{k}\cdot(l-m)}, \tag{8}$$

where $\beta=1/T$, the inverse of temperature. We have set Boltzman constant $k_B=1$. In Eq. (8), let $t = t'$ and $l = m$, then under the RPA, the expression of the magnetization can be solved from an ordinary differential equation of the second order [4,19,28,34,35]:

$$\langle S^z\rangle = \frac{(\Phi + 1 + S)\Phi^{2S+1} - (\Phi - S)(\Phi + 1)^{2S+1}}{(\Phi + 1)^{2S+1} - \Phi^{2S+1}}, \tag{9}$$

where

$$\Phi = \sum_k \frac{1}{e^{\beta\omega(\mathbf{k})} - 1}. \tag{10}$$

From Eqs. (9) and (16), $\langle S^z\rangle$ is computed iteratively. Consequently, the following three correlations can be evaluated:

$$\langle (S^z)^2\rangle = S_p - (1 + 2\Phi)\langle S^z\rangle, \tag{11}$$

$$\langle (S^z)^3\rangle = [(1 + 2\Phi)[S_p - 3\langle (S^z)^2\rangle] + (2S_p - 1)\langle S^z\rangle]/2 \tag{12}$$

and

$$\langle (S^z)^4\rangle = S_p^2 - \langle (S^z)^2\rangle - 2(1 + 2\Phi)\langle (S^z)^3\rangle. \tag{13}$$

From Eq. (8) two formulas can be derived. Note the definition of the operators $A$ and $B$. Substituting them into (8), letting $t = t'$, taking derivative $n$ times with respective to the parameter $u$, letting $u=0$ and then summing over the nearest neighbors of site $m$, we obtain

$$J \sum_j \langle (S_m^z)^n S_m^- S_j^+ \rangle = \Phi_a \langle [S^+, (S^z)^n S^-] \rangle, \qquad (14)$$

where we have defined

$$\Phi_a = \frac{1}{N} \sum_k \frac{J(k)}{e^{\beta \omega(k)} - 1}. \qquad (15)$$

Now let us take derivative with respective to time $t$:

$$\langle B_m(u, t')[A_l(t), H] \rangle = \langle [A, B(u)] \rangle \sum_k \frac{\omega(k) e^{-i\omega(k)(t-t')}}{e^{\beta \omega(k)} - 1} e^{-i\mathbf{k} \cdot (\mathbf{l} - \mathbf{m})} \qquad (16)$$

Letting $t = t'$, taking derivative $n$ times with respective to the parameter $u$ and then letting $u=0$, we achieve

$$\langle (S_0^z)^n S_0^- [S_0^+, H] \rangle = \Phi_b \langle [S^+, (S^z)^n S^-] \rangle, \qquad (17)$$

where

$$\Phi_b = \frac{1}{N} \sum_k \frac{\omega(k)}{e^{\beta \omega(k)} - 1}. \qquad (18)$$

Eqs. (14) and (17) are quite useful for calculation of the LCEs.

The transverse correlation energy was defined in Eq. (3), and can be immediately evaluated by taking $n=0$ in Eq. (14):

$$U_{TC} = -\langle S^z \rangle \Phi_a. \qquad (19)$$

This result was in fact already available in text books, and is a precise one.

It is not so easy to put down a precise expression for the LCE ULC. Some approximations are inevitable.

Multiplying on Eq. (10) an operator $S_i^-$ to the left leads to

$$S_i^-[S_i^+, H] = -J S_i^- \sum_j (S_i^z S_j^+ - S_i^+ S_j^z) + K_z S_i^-(S_i^z S_i^+ + S_i^+ S_i^z) + b_z S_i^- S_i^+. \qquad (20)$$

Reordering the terms results in

$$J \sum_j \langle S_i^z S_j^z \rangle = -\langle S_i^-[S_i^+, H] \rangle + K_z \langle S_i^-(S_i^z S_i^+ + S_i^+ S_i^z) \rangle + b_z \langle S_i^- S_i^+ \rangle$$
$$- J \sum_j \langle S_i^z S_i^- S_j^+ \rangle - J \sum_j \langle S_i^- S_j^+ \rangle + J S_p \sum_j \langle S_j^z \rangle - J \sum_j \langle (S_i^z)^2 S_j^z \rangle. \qquad (21)$$

All the terms except the last one on the right hand side of Eq. (23) can be evaluated by use of Eqs. (14) and (17) without making approximation. For the last term, we have to make approximation as follows

$$\langle (S_i^z)^2 S_j^z \rangle \approx \langle (S_i^z)^2 \rangle \langle S_j^z \rangle. \qquad (22)$$

There are at least three reasons supporting this approximation being a very good one. The first reason is that $\langle (S_i^z)^2 S_j^z \rangle$ is merely one of the terms emerging in the expression of Eq. (21), meaning a small part of $\langle S_i^z S_j^z \rangle$. The second reason is that Eq. (22) is an approximation of a higher order correlation function $\langle (S_0^z)^2 S_j^z \rangle$, which is of cause much better than the decoupling approximation of the lower-order correlation function in Eq. (1). The third reason is that when $S=1/2$, Eq. (22) becomes an identity because $(S^z)^2 = 1/4$, and in this sense, the longitudinal correlation energy can be evaluated precisely.

With the approximation Eq. (22), we are able to put down the LCE

$$U_{LC} = -\frac{1}{2}[S_p - \langle S^z \rangle - 3\langle (S^z)^2 \rangle]\Phi_a + \langle S^z \rangle \Phi_b - \frac{1}{2}J(0)(S_p - \langle (S^z)^2 \rangle)\langle S^z \rangle$$
$$+ \frac{K_z}{2}[2\langle (S^z)^3 \rangle + 3\langle (S^z)^2 \rangle - (2S_p - 1)\langle S^z \rangle - S_p] - \frac{b_z}{2}[S_p - \langle S^z \rangle - \langle (S^z)^2 \rangle]. \tag{29}$$

When $S=1/2$, we have identities $(S^z)^2 = \frac{1}{4}$ and $(S^z)^2 S^- = -\frac{1}{2} S^z S^-$ [36]. Then Eq. (23) goes back naturally to the form ever obtained[18,28]. In this case the correlation function $\langle S_k^z S_j^z \rangle$ is treated precisely.

Before carrying out the numerical computation, let us discuss the values of the energies at two special temperatures, i.e., $T=0$ and $T=T_C$, in the absence of the external field.

At zero temperature, it is easily seen that $\Phi = \Phi_a = \Phi_b = 0$ and thus $\langle (S^z)^n \rangle = S^n$. Therefore, we have

$$U_{TC}(T=0) = 0 \tag{24}$$

The longitudinal energy is exactly the same as Eq. (5), and the internal energy as well.

At the Curie point $T_C$, $\langle S^z \rangle \to 0$, $\Phi_a = T_C$ and $\Phi_b \to -\frac{T_C}{\langle S^z \rangle}(V_{-1} - 1)$, where $V_{-1} = \frac{1}{N}\sum_k \frac{1}{1-\gamma_{kC}}$. The Curie temperature under the RPA is $T_C = \frac{J(0)S_p}{3V_{-1}}$.

Therefore, one obtains

$$\frac{1}{S_p}U_{TC}(T=T_C) = -\frac{J(0)}{3}(1 - \frac{1}{V_{-1}}) \tag{25}$$

and

$$\frac{1}{S_p}U_{LC}(T=T_C) = \frac{J(0)}{6}(1 - \frac{1}{V_{-1}}), \tag{26}$$

respectively. One feature is that both $U_{TC}/S_p$ and $U_{LC}/S_p$ at the Curie point are independent of the spin quantum number $S$.

It should be noted that $U_{LC}(T_C)$ is positive. This means that near the Curie point, the neighboring spins are mainly antiparallel to each other, although the exchange between them is ferromagnetic. This is an unexpected result. We think this a manifestation of quantum fluctuation along the longitudinal direction. In a quantum magnetic system, there exists both the thermodynamic and quantum fluctuations at finite temperature. The former should be isotropic, while the latter, from this work, seems not. Around the Curie point the transverse correlation energy is negative, which means that the spins are transversely parallel to each other, while it is not along the longitudinal direction.

The LCU $U_{TC}$ is precise, as having been mentioned above, and $U_{TC}(T_C)/S_p$ is independent of $S$. For the longitudinal correlation energy, we have obtained that $U_{LC}(T_C)/S_p$ happen to be also independent of $S$. Considering that the energy expression Eq. (23) is precise for the case of $S=1/2$, this independence is regarded as an evidence that supports the approximation Eq. (22) being very good one.

Anyway, as the transverse correlation effect is strong enough, $U_{TC}(T_C)$ is in magnitude two times of $U_{LC}(T_C)$ so that the sum $U_{TC}(T_C)+U_{LC}(T_C)$ is negative.

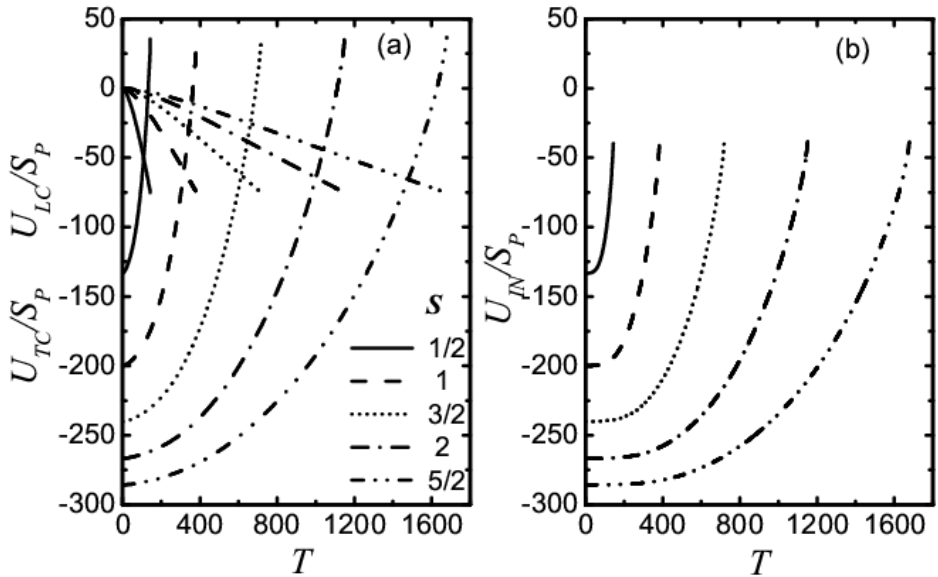

Fig. 1 The energies of bcc FM lattices at $J=100$ for the lowest five $S$ values. (a) The TCEs (descending curves) and LCEs (ascending curves). Near $T_C$, $U_{LC}$'s are positive. (b) The internal energies.

Fig. 1 shows the numerical results. As analyzed above, both $U_{LT}(T_C)/S_p$ and $U_{LC}(T_C)/S_p$ are independent of S, and the latter is indeed greater than zero. Figure 1(a) reveals that the transverse correlation effect becomes stronger as temperature rises, demonstrating that the neighboring spins more likely take antiparallel in the transverse directions. While along the z direction, at T=0, the neighboring spins are strictly parallel to each other. With temperature rising, they deviate from the strict paralleling, and even become antiparallel to each other along this direction when temperature is near the Curie point.

Fig. 2 shows the effect of an external field on the energies in the case of S=1/2. In Fig. 2(a), below the Curie point $U_{TC}$ drops starting from zero temperature as in Fig. 1(a). Obviously, the field depresses the transverse correlation because it forces the spins to be parallel along the z direction. Around the Curie point, this energy turns to increasing with temperature. The turning point slightly rises with the field. Figure 2(b) shows that even if an external field is applied along the z direction, the longitudinal energy around the Curie point is still greater than zero, demonstrating that a magnetic field is unable to totally depress the longitudinal antiparallel correlation between neighboring spins. Above the Curie point, this part of energy is always positive. With the increasing of temperature $U_{LC}(T)/S_p$ gradually decreases toward zero, showing that the correlation fades away with temperature rising. Anyway, the internal energy is always negative and tends to zero as temperature goes to infinite.

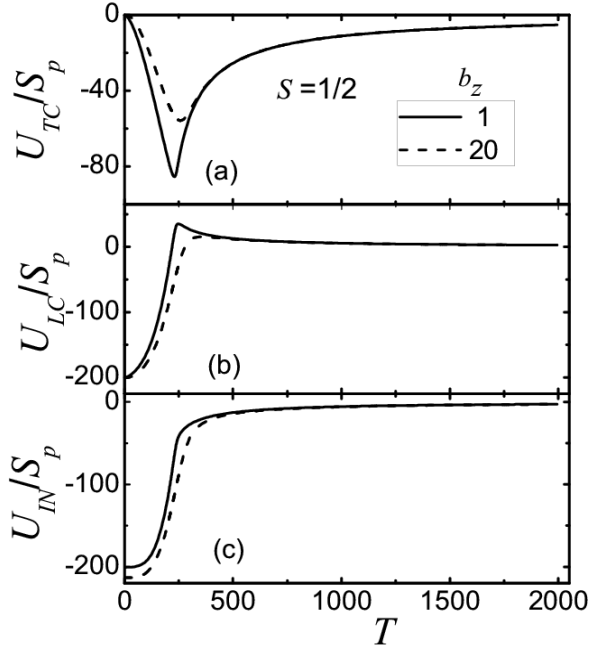

Fig. 2 The energies of a fcc FM lattice for $b_z$=1 and 20 as S=1/2 and J=100. (a) The TCEs. (b) The LCEs. (c) The internal energy. Please note that the Zeeman energy

is not shown.

Equation (20) prompts us that it is possible to evaluate the higher order correlation functions so as to go beyond the approximation Eq. (22). Let us do so. Multiplying $S_i^z$ to Eq. (20) from the left and combining the resultant with Eq. (21) to eliminate the correlation function $\langle (S_i^z)^2 S_j^z \rangle$, we obtain

$$J(S_p+1)\sum_j \langle S_i^z S_j^z \rangle$$
$$= -J\sum_j \langle S_i^- S_j^+ \rangle + J\sum_j \langle (S_i^z)^2 S_i^- S_j^+ \rangle - \langle S_i^-[S_i^+, H] \rangle + \langle S_i^z S_i^-[S_i^+, H] \rangle \qquad (27)$$
$$+J(0)S_p \langle S^z \rangle - K_z \langle (S_i^z - 1) S_i^- S_i^+ (2S_i^z + 1) \rangle - b_z \langle (S_i^z - 1) S_i^- S_i^+ \rangle + J\sum_j \langle (S_i^z)^3 S_j^z \rangle$$

Again, all the terms on the right hand side except the last one can be evaluated by Eqs. (14) and (17). This time, the necessary approximation is

$$\langle (S_i^z)^3 S_j^z \rangle \approx \langle (S_i^z)^3 \rangle \langle S_j^z \rangle . \qquad (28)$$

This approximation is made for the correlation function that is one order higher than that in Eq. (22). Consequently, the achieved LCE is

$$U_{LC}(S>1) = -\frac{1}{2(S_p+1)}\{[S_p - (2S_p+1)\langle S^z \rangle - 3\langle (S^z)^2 \rangle + 4\langle (S^z)^3 \rangle]\Phi_a$$
$$+[-S_p - 3\langle S^z \rangle + 3\langle (S^z)^2 \rangle]\Phi_b + J(0)(\langle (S^z)^3 \rangle + S_p)\langle S^z \rangle \qquad (29)$$
$$+K_2[2\langle (S^z)^4 \rangle + \langle (S^z)^3 \rangle - 2(S_p+1)\langle (S^z)^2 \rangle + (S_p-1)\langle S^z \rangle + S_p]$$
$$+b_z[S_p - (S_p+1)\langle S^z \rangle + \langle (S^z)^3 \rangle]\}.$$

This expression is no doubt better than Eq. (29).

In the case of $S=1$, one should be cautious in dealing with Eq. (27). He should not only employ the identity $(S^z)^3 = S^z$, but also make reduction of the operator production $(S_i^z)^2 S_i^-$ [36]. After these manipulations, one will gain

$$U_{LC}(S=1) = -\frac{1}{4}[-2 - \langle S^z \rangle + \langle (S^z)^2 \rangle]\Phi_1 - \frac{1}{4}[2 - \langle S^z \rangle - 3\langle (S^z)^2 \rangle]\Phi_2$$
$$-\frac{1}{2}J(0)\langle S^z \rangle - \frac{1}{2}K_z[-2\langle (S^z)^2 \rangle + \langle S^z \rangle + 1] - \frac{1}{2}b_z(1 - \langle S^z \rangle). \qquad (30)$$

This expression is precise, because no approximation is made in deriving it. We can also deem that the correlation function $\langle (S_i^z)^2 S_j^z \rangle$ is treated precisely.

Once more, we reckon the values at zero temperature and the Curie point when the field is absent. At zero temperature, both Eqs. (38) and (39) reach

$U_{LC} = -\frac{1}{2}J(0)S^2$. At the Curie point, the energy values can be written as

$$\frac{1}{S_p}U_{LC}(S > \frac{1}{2}, T_C) = g(S)\frac{J(0)}{6}(1 - \frac{1}{V_{-1}}) + K_{2a}(S) \tag{31}$$

where $g(S=1) = \frac{1}{2}$, $K_{2a}(S=1) = \frac{1}{12}K_2$, $g(S>1) = \frac{-2S_p + 9}{5(S_p + 1)}$ and $K_{2a}(S>1) = \frac{K_2(4S_p - 3)}{30(S_p + 1)}$. Firstly, let us see the case where the anisotropy is absent.

The factor $g(S)$ demonstrates that the quantity $U_{LC}(T_C)/S_p$ is now dependent on $S$. It is noticed that as $S = 3/2$, $g(3/2) = 6/95$, and $g(S>3/2)<0$. As $S$ goes to infinity, the factor $g(S)$ approaches $-2/5$. This discloses that for $S>3/2$, the neighboring spins are not antiparallel. Only for $S<3/2$, will the neighboring spins be antiparallel to each other around the Curie point. This shows that the longitudinal correlation effect is actually closely related to the spin quantum number $S$. A system with the lower $S$ has the stronger quantum fluctuation. Secondly, when there is an single-ion anisotropy, the $K_{2a}(S)$ term in Eq. (31) is always positive, i.e., this kind of anisotropy strengthens the longitudinal quantum fluctuation. By contrast, this anisotropy did not play a role in Eq. (26).

Fig. 3 presents the numerical results. The lines of $S=1/2$ are just those in Fig. 1. The $S=1$ lines are calculated from Eq. (30), and remaining lines from Eq. (29). Figure 3(a) demonstrates that the quantum fluctuation decays with spin quantum number. The smallest $S$ has the strongest quantum fluctuation.

In Ref. [18], the energy expression of $S=1$ was given when a higher-order Green's function $\langle\langle S_i^+; S_j^z S_j^-\rangle\rangle$ was employed. We retrieved their derivation, and the numerical results revealed that the TCE and LCE are exactly the same as the dashed curves in Fig. 3. This discloses that the precise process of $\langle (S_i^z)^2 S_j^z \rangle$ is equivalent to the inclusion of the higher order Green's function $\langle\langle S_i^+; S_j^z S_j^-\rangle\rangle$.

Fig. 4 shows the effect of an external field on the energies in the case of $S=5/2$. The overall behavior of the curves are the same as those in Fig. 2, except that the longitudinal correlation energies in the case of $S=5/2$ are always negative, as shown in Fig. 4(b).

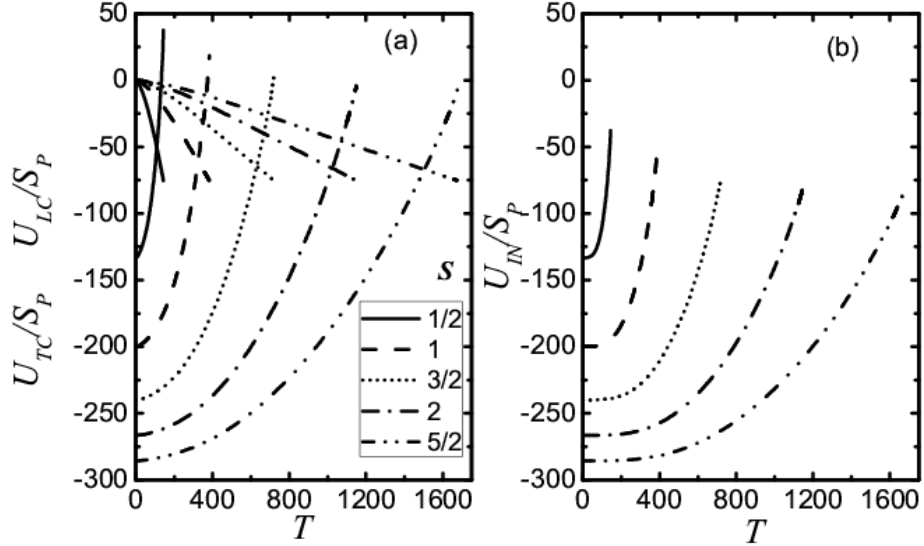

Fig. 3 The energies of bcc FM lattices at $J=100$ for the lowest five $S$ values by Eq. (29) for $S=1/2$, Eq. (39) for $S=1$ and Eqs. (38) for $S>1$, where the anisotropy and field are absent. (a) The TCE (descending curves) and LCEs (ascending curves). The former are the same as those in Fig. 1(a). Up to $T_C$, $U_{LC}$ remains negative for $S>3/2$. (b) The internal energies.

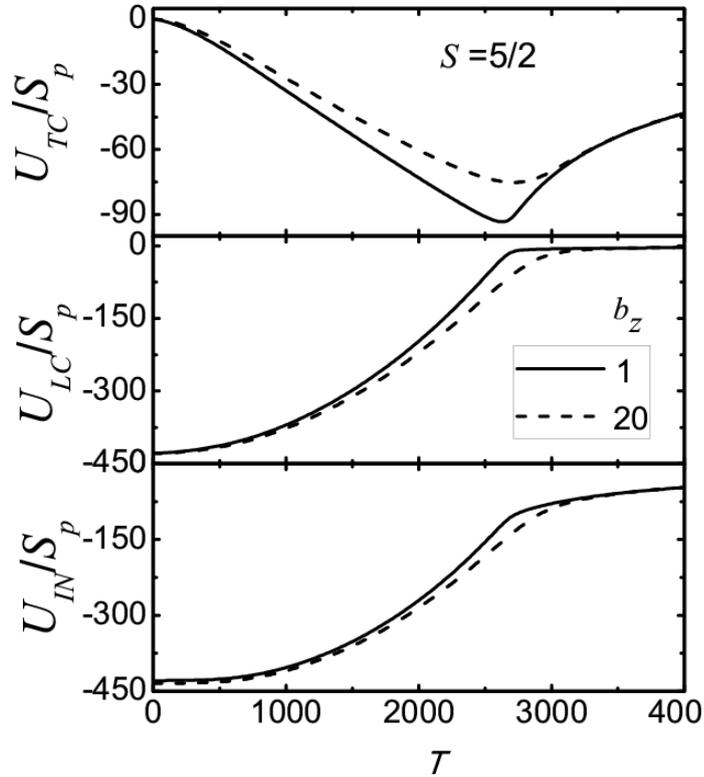

Fig. 4 The energies of a fcc FM lattice for $b_z=1$ and 20 as $S=5/2$ and $J=100$. (a) The TCEs. (b) The LCEs. (c) The internal energy. Please note that the Zeeman energy is not shown.

The energy formulas Eqs. (29) and (30) stand for the case where the magnetization is nonzero. It was pointed out that the influence of quantum spin effects on the magnetic short-range order increased with a decrease in quantum spin value [27]. Our work discloses that in the presence of the magnetic long-range order, a smaller spin quantum number has a stronger longitudinal quantum spin effect.

### III. One-component magnetization: Antiferromagnetic systems

The Hamiltonian is now

$$H = -\frac{1}{2} J \sum_{i,j} \langle S_{1i}^+ S_{2j}^- + S_{1i}^z S_{2j}^z \rangle - \sum_{\mu=1}^{2} [K_{z\mu} \sum_i \langle (S_{\mu i}^z)^2 \rangle + b_z \sum_i \langle S_{\mu i}^z \rangle] \qquad (32)$$

Here $J<0$. We assume that the lattice is divided into two sublattices, each with $N/2$ sites, which are labeled by the lower case Greek letters such as $\mu$, $\mu=1, 2$. The spin quantum numbers might not be the same, denoted as $S_1$ and $S_2$, respectively, so that the system might be a ferrimagnetic one. Correspondingly, we define $S_{p\mu} = S_\mu(S_\mu+1), \mu=1,2$.

It should be emphasized that the following treatment of the AFM systems is not applicable to the case of zero temperature. Hence, hereafter, when we mention $T=0$, we actually mean the temperature very close to zero.

Now we choose the operators $A = (S_{1m}^+, S_{2n}^+)^{\mathrm{T}}$, $B = (\exp(uS_{1i}^z)S_{1i}^-, \exp(uS_{2j}^z)S_{2j}^-)$

to construct Green's function $G(t,t') = \langle\langle A(t); B(t') \rangle\rangle$ This is a matrix, and its Fourier component is denoted as $g(\omega)$. The application of the EOM leads to a linear equations:

$$[\omega I - P]g = F_{-1}. \qquad (33)$$

where $F_{-1}$ is the commutator matrix of operators defined by $F_{-1} = \langle [A,B] \rangle$. The matrix $P$ is

$$P = \begin{pmatrix} J(0)\langle S_2^z \rangle + K_{z1}C_1\langle S_1^z \rangle + b_z & -J(k)\langle S_1^z \rangle \\ -J(k)\langle S_2^z \rangle & J(0)\langle S_1^z \rangle + K_{z2}C_2\langle S_2^z \rangle + b_z \end{pmatrix} \qquad (34)$$

Its eigenvalues $\omega_\tau(k)$, $\tau=1,2$ can be solved, and the eigenvector matrix $U$ and its inverse $U^{-1}$ of $P$ as well.

Although the Green's functions are in a matrix form, we can follow the routine almost the same as the last section so as to obtain the internal energy of the system. The TCE and LCE are

$$U_{TC} = -\frac{1}{2} J \sum_{n} \langle S_{2n}^- S_{1m}^+ \rangle = \Phi_{a,21} \langle S_1^z \rangle \qquad (35)$$

and

$$\begin{aligned}U_{LC} &= -\frac{1}{2} J \sum_{n} \langle S_{1m}^z S_{2n}^z \rangle \\ &= \frac{1}{2}\{[-S_{p1} + \langle S_1^z \rangle + 3\langle (S_1^z)^2 \rangle]\Phi_{a,21} + 2\langle S_1^z \rangle \Phi_{b,1} - J(0)[S_{p1} - \langle (S_1^z)^2 \rangle]\langle S_2^z \rangle \\ &\quad + K_{z1}[2\langle (S_1^z)^3 \rangle + 3\langle (S_1^z)^2 \rangle - (2S_{p1}-1)\langle S_1^z \rangle - S_{p1}] - b_z[S_{p1} - \langle S_1^z \rangle - \langle (S_1^z)^2 \rangle]\}\end{aligned} \qquad (36)$$

respectively, where we have defined the notations

$$\Phi_{a,\mu\nu} = \frac{2}{N} \sum_{k} \sum_{\tau} \frac{U_{\mu\tau} U_{\tau\nu}^{-1}}{e^{\beta\omega_\tau} - 1} J(k) \qquad (37)$$

and

$$\Phi_{b,\mu} = \frac{2}{N} \sum_{k} \sum_{\tau} \frac{\omega_\tau U_{\mu\tau} U_{\tau\mu}^{-1}}{e^{\beta\omega_\tau} - 1}. \qquad (38)$$

The results obtained in Ref. [29] for $S=1/2$ can be retrieved in the same way with the caution that the exchange was anisotropic.

For an AFM lattice, $S_1 = S_2 = S$. When the external field is absent, $\langle S_1^z \rangle = -\langle S_2^z \rangle = \langle S^z \rangle$. The energy values at zero temperature and the Néel point can be easily put down. At T=0,

$$\frac{1}{S_p} U_{TC}(T=0) = -\frac{1}{2}(|J(0)| + K_z C) \sum_{k} \frac{\gamma_{kC}^2}{(1-\gamma_{kC}^2)^{1/2}} \qquad (39)$$

and

$$\begin{aligned}U_{LC}(T=0) &= [S_p - \langle S^z \rangle - 3\langle (S^z)^2 \rangle]\Phi_{a,12} + 2\langle S^z \rangle \Phi_{b,1} \\ &\quad - J(0)(\langle (S^z)^2 \rangle - S_p)\langle S^z \rangle - K_z[\langle -2(S^z)^3 - 3(S^z)^2 + (2S_p-1)S^z \rangle + S_p].\end{aligned} \qquad (40)$$

At $T=T_N$,

$$\frac{1}{S_p} U_{TC}(T_N) = -\frac{J(0)}{3}(\frac{1}{V_{-1}} - 1) \qquad (41)$$

and

$$\frac{1}{S_p} U_{LC}(T_N) = \frac{J(0)}{6}(\frac{1}{V_{-1}} - 1). \qquad (42)$$

The right hand sides of Eqs. (41) and (42) are in fact exactly the same as that of Eqs. (25) and (26), which is due to the fact that under RPA the Néel point $T_N$ of the antiferromagnet is the same as the Curie point of the ferromagnet with the same exchange strength. Again, the longitudinal correlation energy is positive around the transition temperature, which means that the neighboring spins are parallel to each other, although the exchange between them is antiferromagnetic.

Fig. 5 shows the transverse and longitudinal correlation energies of bcc AFM lattices without the field the anisotropy.

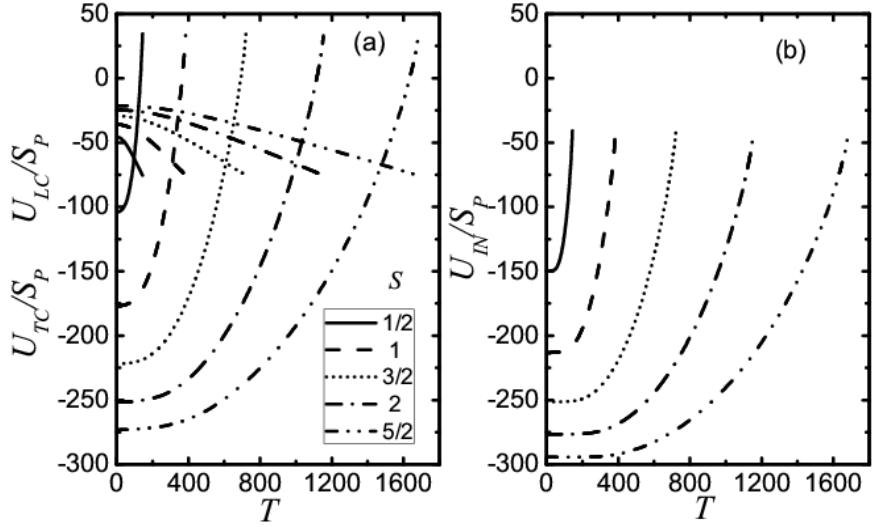

Fig. 5 The energies of bcc AFM lattices when the anisotropy and external field are absent at $J= -100$ for the lowest five $S$ values. (a) The TCEs (descending curves) and LCEs (ascending curves). Near $T_C$, $U_{LC}$'s are positive. (b) The internal energies.

We proceed to deal with the correlation function $\langle (S_i^z)^2 S_j^z \rangle$ in the way similar to Eq. (27) and to take the approximation of the higher order correlation function as Eq. (28). The resultant longitudinal correlation energies are

$$U_{LC}(S>1) = -\frac{1}{S_{p1}+1}\{[S_{p1} - (2S_{p1}+1)\langle S_1^z\rangle - 3\langle (S_1^z)^2\rangle + 4\langle (S_1^z)^3\rangle]\Phi_{a,21}$$
$$+[-S_{p1} - 3\langle S_1^z\rangle + 3\langle (S_1^z)^2\rangle]\Phi_{b,1} + J(0)\langle (S_1^z)^3\rangle\langle S_2^z\rangle + J(0)S_{p1}\langle S_2^z\rangle \quad (43)$$
$$+K_{z1}\langle 2\langle (S_1^z)^4\rangle + \langle (S_1^z)^3\rangle - 2(S_{p1}+1)\langle (S_1^z)^2\rangle + (S_{p1}-1)\langle S_1^z\rangle + S_{p1}\rangle$$
$$+b_z[S_{p1} - (S_{p1}+1)\langle S_1^z\rangle + \langle (S_1^z)^3\rangle]\}$$

and

$$U_{LC}(S=1) = -\frac{1}{4}[2 - \langle S_1^z\rangle - 3\langle (S_1^z)^2\rangle]\Phi_{a,21} - \frac{1}{4}[-2 - 3\langle S_1^z\rangle + 3\langle (S_1^z)^2\rangle]\Phi_{b,1}$$
$$-\frac{1}{2}J(0)\langle S_2^z\rangle + \frac{1}{2}K_{z1}[2\langle (S_1^z)^2\rangle - \langle S_1^z\rangle - 1] + \frac{1}{2}b_z(\langle S_1^z\rangle - 1). \quad (44)$$

When the external field is absent, the longitudinal energies of an antiferromagnet at zero temperature and the Néel point are

$$\frac{1}{S_p}U_{LC}(T=0,S>1)$$

$$=-\frac{1}{S_p(S_p+1)}\{[S_{p1}-(2S_{p1}+1)\langle S^z\rangle-3\langle(S^z)^2\rangle+4\langle(S^z)^3\rangle]\Phi_{a,21} \quad (45)$$

$$+[-S_p-3\langle S^z\rangle+3\langle(S^z)^2\rangle]\Phi_{b,1}-J(0)\langle(S^z)^3\rangle\langle S^z\rangle-J(0)S_p\langle S^z\rangle$$

$$+K_z\langle 2\langle(S^z)^4\rangle+\langle(S^z)^3\rangle-2(S_p+1)\langle(S^z)^2\rangle+(S_p-1)\langle S^z\rangle+S_p\rangle\},$$

$$\frac{1}{2}U_{LC}(T=0,S=1)=\frac{1}{2}\{-\frac{1}{4}[2-\langle S^z\rangle-3\langle(S^z)^2\rangle]\Phi_{a,21}$$

$$-\frac{1}{4}[-2-3\langle S^z\rangle+3\langle(S^z)^2\rangle]\Phi_{b,1}+\frac{1}{2}J(0)\langle S^z\rangle+\frac{1}{2}K_z[2\langle(S^z)^2\rangle-\langle S^z\rangle-1]\} \quad (46)$$

and

$$\frac{1}{S_p}U_{LC}(S>\frac{1}{2},T_N)=g(S)\frac{J(0)}{6}(1-\frac{1}{V_{-1}})+K_{2a}(S), \quad (47)$$

where $g(S)$ and $K_{2a}(S)$ are the same as those in Eq. (40).

    We do not plot the curves of the longitudinal correlation energies versus temperature presented by Eq. (43) and (44). We merely mention two facts from which one can be aware of the curves. One is that at the transition point, the right hand side of Eq. (47) is exactly the same as Eq. (31). The other is the comparison of Eqs. (45) (46) with (40) at $T=0$. It is difficult to prove that they are equal to each other, but the numerical results turn out that they do not have significant difference. This demonstrates that the consideration of the higher longitudinal correlation functions significantly improves the longitudinal correlation energies near $T=T_N$, but does not so near T=0.

## IV. Three-component magnetization for ferromagnetic systems

    In the above two sections, both the anisotropy and the field point to the z direction, so that the magnetization does. If the field orients arbitrarily, the Hamiltonian should be written, instead of Eq. (2), as

$$H=-\frac{1}{2}J\sum_{i,j}\mathbf{S}_i\cdot\mathbf{S}_j-K_z\sum_i(S_i^z)^2-\mathbf{b}\cdot\sum_i\mathbf{S}_i. \quad (48)$$

Subsequently, the magnetization may have more than one component. This case was first studied by Fröbrich et al. [12,13] for FM films. Following their work, we investigated the bulk systems and gave a formula for evaluating three-component magnetizations applicable to any $S$ values [16]. The calculation of the internal energy in this case has not been touched yet.

    The internal energy should be

$$U_{IN} = -\frac{1}{2}J\sum_j \langle S_i^+ S_j^- \rangle - \frac{1}{2}J\sum_j \langle S_i^z S_j^z \rangle - K_z \langle (S^z)^2 \rangle - \boldsymbol{b}\cdot\langle \boldsymbol{S}\rangle \quad (49)$$

As long as the magnetization is nonzero, its z-component is so due to the existence of the $K_z$ term. In spite of this fact, the first two terms in Eq. (49) should not be regarded as the transverse and longitudinal correlation energies, because the word "longitudinal" refers to the direction parallel to the magnetization, not just one of its components. They are therefore denoted instead as $U_{xy}$ and $U_z$, respectively.

As has been presented in Refs. [16] and [34], we have to choose the operators $\boldsymbol{A} = (S^+, S^-, S^z)^T$, $\boldsymbol{B} = \exp(uS^z)(S^+, S^-, S^z)$ to construct the Green's function $\boldsymbol{G}_{ij}(t,t') = \langle\langle \boldsymbol{A}_i; \boldsymbol{B}_j \rangle\rangle$. This is a matrix Green's function, and its Fourier component is denoted as $\boldsymbol{g}(\omega)$. The application of the EOM leads to a linear equations in the form of Eq. (33). with the matrix $\boldsymbol{F}_{-1}$ being $\boldsymbol{F}_{-1} = \langle [\boldsymbol{A}_i, \boldsymbol{B}_j] \rangle$, and the matrix $\boldsymbol{P}$ and the solution of the linear equations were presented in Ref. [16]. The following quantities were essential: $E_k = \sqrt{H_+ H_- + H_z^2}$, where $H_z = J(0)(1-\gamma_k)\langle S^z \rangle + K_z C \langle S^z \rangle + b_z$, $H_\pm = J(0)(1-\gamma_k)\langle S^\pm \rangle + b_\pm$, $\gamma_k = J(\boldsymbol{k})/J(0)$ and $b_\pm = b_x \pm ib_y$.

Under the RPA, the magnetization components observe a regularity condition[12,13,16], from which one obtains

$$q_\alpha \langle S^\beta \rangle = q_\beta \langle S^\alpha \rangle, \alpha = +,-,z, \quad (50)$$

where

$$q_\alpha(\boldsymbol{k}) = \frac{H_\alpha(\boldsymbol{k})}{E(\boldsymbol{k})\coth(\beta E(\boldsymbol{k})/2)}, \alpha = +,-,z. \quad (51)$$

In terms of the spectral theorem, the expression of the $U_{xy}$ can be derived:

$$U_{xy} = -\frac{1}{4}(5 - \frac{3}{R^2})\langle S^z \rangle \sum_k \frac{J(\boldsymbol{k})}{q_z} \quad (52)$$

where

$$R^2 = 1 + |q_+/q_z|^2. \quad (53)$$

When the field just point to the z axis, and the magnetization is along this direction without other components, we have $H_\pm = 0$ and $q_\pm = 0$. Then it can be checked that Eq. (52) goes back to Eq. (19). In this case, the $U_{xy}$ can be regarded as the TCE.

The merit of evaluation of the three-component magnetization is that all the necessary correlation functions, including $\langle S_i^z S_j^z \rangle$, can be evaluated. This enables one to calculate the $U_z$ without resorting to Eqs. (14) and (17). Here we simply put down the final result without giving the tedious derivation:

$$U_z = -\frac{1}{4}(1-\frac{3}{R^2})\langle S^z \rangle \sum_k \frac{J(\boldsymbol{k})}{q_z}. \tag{54}$$

The energies being in this form, it is difficult to estimate theirs values at transition temperature as we did in the last two sections. Nevertheless, it is guaranteed that $U_{xy}+U_z$ is always negative.

It is our reluctance to point the two shortcomings of Eq. (54). This equation is not valid for $q_\pm = 0$, since in the course of deriving it we have divided $q_+$. Thus, unlike the case in Sec. II, where the energies can be calculated even when the magnetization tends to be zero, Eq. (54) stands for the cases when $q_+$, as well as $\langle S^+ \rangle$, does not tend to be zero. This is one of the shortcomings of Eq. (54). In Sec. II, one is able to explore the better expression of the LCE Eqs. (29) and (30) by making the approximation to the higher order correlation functions such as Eq. (28). The key is to employ Eqs. (14) and (17). In the present section, these equations are not used, and the longitudinal correlation function $\langle S_i^z S_j^z \rangle$ is directly calculated within the RPA. Therefore, the degree of approximation of Eq. (54) should equivalent to that of Eq. (23), and it is difficult to explore the better expressions like Eq. (29). This is the other shortcoming of Eq. (54).

## V. Summary

In this paper, we derived the internal energies of some magnetic system modeled by Heisenberg Hamiltonian with the nearest neighbor exchanges. The internal energy mainly include two parts: the transverse and longitudinal correlation energies. We first derive the expressions for FM systems where the magnetizations are along the $z$ direction. By making use of Eqs. (14) and (17) derived from the spectral theorem Eq. (8), we are able to reckon higher order longitudinal correlation functions such as $\langle (S_i^z)^2 S_j^z \rangle$ so as to achieve better results, e. g., Eqs. (29) and (30). For AFM systems, the deriving procedure is similar to that of FM ones. An interesting result is that for the three smallest spin quantum numbers, the longitudinal correlation energies around the transition temperatures are positive for both FM and AFM systems, which means that the neighboring spins in FM (AFM) systems are antiparallel (parallel) to each other in spite of the FM (AFM) exchanges between them. This is attributed to quantum fluctuation which is believed anisotropic. A smaller spin has a stronger quantum longitudinal fluctuation, and this effect cannot be totally suppressed by an external magnetic field. The consideration of the higher longitudinal correlation functions significantly improves the longitudinal correlation energies near the transition temperatures for both FM and AFM systems, while it does not so near $T$=0 for AFM ones.

At last, the case of three-component magnetization of FM systems is investigated. The main parts of the energy are $U_{xy}$ and $U_z$. When the magnetization point to the z direction, the expression of $U_{xy}$ becomes the same as that of TCE. The two shortcomings of the expression of $U_z$ are pointed out.

Since now the expressions of the internal energy have been available, see Eqs. (29), (30), (43) and (44), other thermodynamic quantities such as free energy can be calculated consequently.

## ACKNOWLEDGMENTS

This work is supported by the 973 Project of China (Grant No. 2012CB927402) and the National Natural Science Foundation of China (Grant Nos. 11074145 and 61275028).